\documentclass[aps,pra,superscriptaddress,twocolumn,showpacs,longbibliography]{revtex4-1}

\usepackage{amssymb}
\usepackage{amsmath}
\usepackage{amsthm}
\usepackage{graphicx}
\usepackage{amsfonts}
\usepackage{color}
\usepackage{times}
\usepackage{natbib}
\usepackage{comment}
\usepackage{soul}
\usepackage{booktabs}
\usepackage{pifont}
\usepackage{hyperref}
\hypersetup{
        colorlinks=true,
}

\newcommand{\beq}{\begin{equation}}
\newcommand{\eeq}{\end{equation}}
\newcommand{\bea}{\begin{eqnarray}}
\newcommand{\eea}{\end{eqnarray}}
\newcommand{\ba}{\begin{align}}
\newcommand{\ea}{\end{align}}

\renewcommand{\AA}

\begin{document}
\title{
Autonomous quantum refrigerator in a circuit-QED architecture based on a Josephson junction}

\author{Patrick P. Hofer}\email{patrick.hofer@unige.ch}\affiliation{D\'epartement de Physique Th\'eorique, Universit\'e de Gen\`eve, 1211 Gen\`eve, Switzerland}
\author{Mart\' i Perarnau-Llobet}\affiliation{D\'epartement de Physique Th\'eorique, Universit\'e de Gen\`eve, 1211 Gen\`eve, Switzerland}\affiliation{ICFO-Institut de Ciencies Fotoniques, The Barcelona Institute of Science and Technology, 08860 Castelldefels, Barcelona, Spain}
\author{Jonatan Bohr Brask}\affiliation{Group of Applied Physics, University of Geneva, Switzerland}
\author{Ralph Silva}\affiliation{D\'epartement de Physique Th\'eorique, Universit\'e de Gen\`eve, 1211 Gen\`eve, Switzerland}
\author{Marcus Huber}\affiliation{Institute for Quantum Optics and Quantum Information (IQOQI), Austrian Academy of Sciences, Boltzmanngasse 3, A-1090 Vienna, Austria}
\affiliation{Group of Applied Physics, University of Geneva, Switzerland}
\author{Nicolas Brunner}\affiliation{D\'epartement de Physique Th\'eorique, Universit\'e de Gen\`eve, 1211 Gen\`eve, Switzerland}

\begin{abstract}
An implementation of a small quantum absorption refrigerator in a circuit QED architecture is proposed. The setup consists of three harmonic oscillators coupled to a Josephson junction. The refrigerator is autonomous in the sense that it does not require any external control for cooling, but only thermal contact between the oscillators and heat baths at different temperatures. In addition, the setup features a built-in switch, which allows the cooling to be turned on and off. If timing control is available, this enables the possibility for coherence-enhanced cooling. Finally, we show that significant cooling can be achieved with experimentally realistic parameters and that our setup should be within reach of current technology.
\end{abstract}

\maketitle

\section{Introduction}
Thermodynamics emerged with the advent of heat machines from the need to understand their fundamental limitations of transforming energy. Rapid advance in quantum technologies have sparked renewed interest into quantum scale machines and their governing thermodynamic laws. The fact that the fundamental carriers of energy exhibit a much richer set of possible states, due to quantum superposition, coherence and entanglement, has fueled a growing interest from the community of quantum information \cite{goold:2015,vinjanampathy:2016}.  A prominent direction for exploring these exciting new possibilities is to study small quantum machines made up of a few quantum systems.

At it's heart thermodynamics has always been a resource theory -- it was conceived out of the need to understand which are the relevant resources that allow for the transformation of energy in desireable ways. From running the first steam engines to achieving ultra-cold temperatures in modern experiments, the quest is arguably still the identification of the best ways to achieve this with the means at our disposal. What changes when transferring these questions to the quantum realm, is what we perceive as an easy task or a free resource. Thermalization, that is the equilibration towards thermal equilibrium, is still an ubiquitous phenomenon in quantum systems \cite{gogolin:2016}, which justifies the assumption that thermal states are free resources \cite{brandao:2013}, just as in classical systems. One fundamental difference, however, is the cost of observation and control. For classical machines the influence of observation of the internal dynamics is negligible and implementing controlled cycles (depending on precise timing or observation) can essentially be neglected, as it adds only very little extra cost. 
The situation is clearly completely different for quantum machines.
For instance, quantum machines working under externally controlled cycles, such as recently reported implementations \cite{rossnagel:2016}, require additional energy to be operated. The cost of this classical control (i.e. for engineering time-dependent Hamiltonians) is typically many orders of magnitude higher than the energy generated by the quantum process, which makes it challenging to design efficient thermal machines.

A promising alternative comes from the exploration of autonomous quantum thermal machines, see e.g. \cite{linden:2010prl,palao:2001,levy:2012,brunner:2012,silva:2015} and \cite{goold:2015,kosloff:2014} for recent reviews. These operate purely on access to the surrounding thermal baths and only require access to two different temperatures, hence avoiding any external control. A paradigmatic example is the quantum absorption refrigerator \cite{linden:2010prl,levy:2012}, which uses a thermal gradient to directly cool down a quantum system. It has been shown that genuine quantum features, such as entanglement \cite{brunner:2014} or coherence \cite{uzdin:2015} may improve the operational range of these fridges. In particular it has been found that the finite-time behaviour, the so called transient regime, leads to genuine quantum effects and the potential to reach even lower temperatures compared to the steady-state regime \cite{mitchison:2015,brask:2015}. While these theoretical developments offer exciting prospects for autonomous quantum thermal machines, their experimental implementation remains mostly unexplored, although some proposals were reported considering various physical systems \cite{chen:2012,venturelli:2013,mari:2012,leggio:2015,brask:2015njp,mitchison:2016,hofer:2016prb}. Moreover, thermoelectric devices also provide a natural platform to study autonomous thermal machines, see e.g. \cite{sothmann:2015,thierschmann:2015,bergenfeldt:2014,hofer:2015}.

Here we propose an implementation of a quantum absorption refrigerator in a circuit QED architecture. Specifically, we consider a setup consisting of three harmonic oscillators, provided by microwave cavities, coupled to a Josephson junction. Other proposals that make use of similar architectures include a quantum heat engine \cite{hofer:2016prb}, a Fock-state stabilizer \cite{souquet:2016}, and a study on Majorana zero modes \cite{dmytruk:2016}, indicating the versatility of such setups. 

The dynamics in our system can be engineered such that the system operates as an autonomous refrigerator. Moreover, we show that the setup features a built-in ``on/off-switch'' which allows one to control the operation mode, i.e. the possibility of turning the fridge on and off. This on/off-switch allows for coherence-enhanced cooling in the transient regime, whenever timing control is available. Finally, we discuss the prospects of an experimental implementation of our setup, which appears promising with current day technology.

\begin{figure*}[t!]
  \centering
  \includegraphics[width=.85\textwidth]{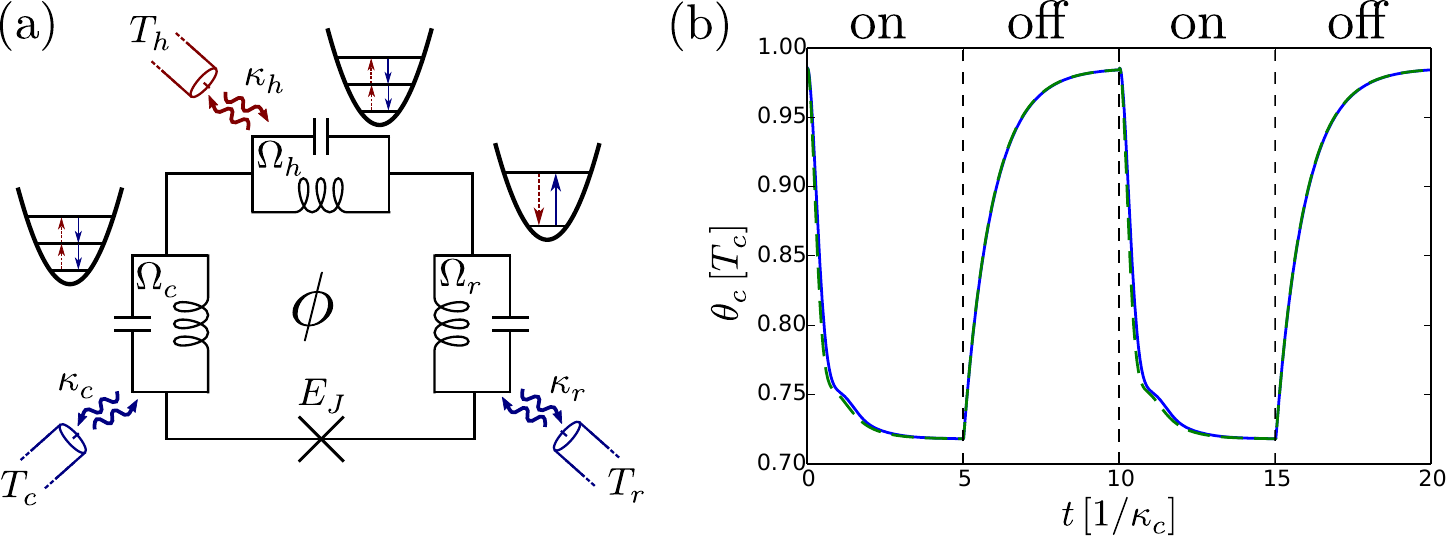}
  \caption{(a) Sketch of the refrigerator. A josephson junction is coupled to three $LC$-circuits acting as harmonic oscillators which are themselves coupled to thermal baths. The phase across the junction is determined by the oscillators as well as a magnetic flux through the loop structure. The three-body interaction mediated by the Josephson junction allows for two photons with energies $\Omega_c$ and $\Omega_h$ to be converted into a photon with energy $\Omega_r=\Omega_c+\Omega_h$. A hot bath at temperature $T_h$ ensures an elevated occupation number in oscillator $h$ favoring the process illustrated with blue (solid) arrows over the process illustrated with red (dashed) arrows. (b) On/off refrigeration. Plot of the temperature in the $c$ oscillator $\theta_c$ obtained from the energy of the reduced density matrix (see main text). At the dashed vertical lines, the refrigerator is switched on or off respectively allowing for on-demand cooling of a harmonic oscillator. As discussed in App.~\ref{app:state}, the reduced state of the $c$ oscillator is very close to a thermal state. The green dashed line is obtained with the simplified model in Eqs.~\eqref{eq:toyon} and \eqref{eq:toyoff} with $E_J'/E_J=0.059$ and $E_J''/E_J=0.00135$.}
  \label{fig:sketch}
\end{figure*}

\section{Model}
Our system is sketched in Fig.~\ref{fig:sketch} (a). It consists of three harmonic oscillators provided by $LC$-resonators coupled to a Josephson junction. Each harmonic oscillator is coupled to a heat bath with respective temperatures $T_c$ (cold), $T_h$ (hot), and $T_r$ (room) with $T_c \leq T_r < T_h$. We will use the subscripts $c$, $h$, and $r$ to denote the quantities associated with the harmonic oscillator coupled to the respective bath. In addition to the harmonic oscillators, the Josephson junction is externally phase biased which can be realized in a loop geometry with a magnetic field as sketched in Fig.~\ref{fig:sketch} (a).

This system (without baths) is described by the following Hamiltonian \cite{armour:2013,gramich:2013,trif:2015}
\begin{equation}
\label{eq:hamiltonian}
\hat{H}=\sum_{\alpha=h,c,r}\Omega_\alpha\hat{a}^\dag_\alpha\hat{a}_\alpha-E_J\cos(2\hat{\varphi}_c+2\hat{\varphi}_h+2\hat{\varphi}_c+\phi),
\end{equation}
where $\Omega_\alpha$ denotes the frequencies of the harmonic oscillator and $E_J$ is the josephson energy. The phase of the Josephson junction is driven by the flux $\hat{\varphi}_\alpha=\lambda_\alpha(\hat{a}_\alpha+\hat{a}^\dagger_\alpha)$ of each oscillator where $\lambda_\alpha=\sqrt{\pi e^2Z_\alpha/h}$ is determined by the impedance $Z_\alpha$ of resonator $\alpha$. Additionally, the Josephson junction is biased with the external phase $\phi$. 

In order to make the system work as a refrigerator, we will impose the resonance condition 
\begin{equation}
\label{eq:resonance}
\Omega_r=\Omega_c+\Omega_h.
\end{equation}
Next, as discussed in detail in App.~\ref{app:rwa}, we make a rotating wave approximation (RWA) and only keep the most relevant resonant terms in the expansion of the cosine in Eq.~\eqref{eq:hamiltonian}
\begin{subequations}
\begin{align}
&\qquad\hat{H}\simeq\sin\phi\hat{H}_{\rm on}+\cos\phi\hat{H}_{\rm off},\label{eq:honplusoff}\\
\label{eq:hon}
\hat{H}_{\rm on}&=-E_J\left[\hat{a}_r^\dag\hat{A}_h(1)\hat{A}_c(1)\hat{A}_r(1)\hat{a}_c\hat{a}_h+H.c.\right],\\
\label{eq:hoff}
\hat{H}_{\rm off}&=E_J\left[(\hat{a}_r^\dag)^2\hat{A}_h(2)\hat{A}_c(2)\hat{A}_r(2)\hat{a}_c^2\hat{a}_h^2+H.c.\right],
\end{align}
\end{subequations}
where we introduced the Hermitian operators
\begin{equation}
\label{eq:aops}
\hat{A}_\alpha(k)=(2\lambda_\alpha)^ke^{-2\lambda_\alpha^2}\sum\limits_{n_\alpha=0}^{\infty}\frac{n_\alpha!}{(n_\alpha+k)!}L_{n_\alpha}^{(k)}(4\lambda_\alpha^2)|n_\alpha\rangle\langle n_\alpha |,
\end{equation}
with the generalized Laguerre polynomials $L_n^{(k)}(x)$. 

At this point, the cooling mechanism can already be identified. Indeed, the Hamiltonian in Eq.~\eqref{eq:hon} represents a three-body interaction which converts a $c$ and an $h$ photon into an $r$ photon. This process removes a photon from the cold oscillator thereby cooling it. In order to favor this process compared to the reverse process (where a single $r$ photon is converted into a $c$ and an $h$ photon), we will place the system in thermal contact with heat baths. Thus the system operates as a quantum absorption refrigerator, which cools oscillator $c$ by making use of a heat flux from the hot temperature bath to the room temperature bath \cite{brunner:2012}.

Including the coupling to the thermal baths, the dynamics of the system is then described by the master equation
\begin{equation}
\label{eq:master}
\begin{aligned}
\partial_t\hat{\rho}&=-i[\hat{H},\hat{\rho}]\\&+\sum_{\alpha=h,c,r}\left\{\kappa_\alpha(n_B^\alpha+1)\mathcal{D}[\hat{a}_\alpha]\hat{\rho}+\kappa_\alpha n_B^\alpha\mathcal{D}[\hat{a}_\alpha^\dag]\hat{\rho}\right\},
\end{aligned}
\end{equation}
with the standard Lindblad superoperators $\mathcal{D}[\hat{A}]\hat{\rho}=\hat{A}\hat{\rho}\hat{A}^\dag-\{\hat{A}^\dag\hat{A},\hat{\rho}\}/2$, $\kappa_\alpha$ denotes the energy damping rate associated with the bath $\alpha$ and $n_B^\alpha=[\exp({\frac{\Omega_\alpha}{k_BT_\alpha}})-1]^{-1}$ the occupation number of the baths at the relevant frequencies. We note that we make use of a local master equation, where each oscillator only couples to its respective bath. Such an approach is valid (under the usual Born-Markov approximations) as long as $\kappa_\alpha,E_J\ll\Omega_\alpha$ \cite{puri:book}.

As we will now show by numerically solving the master equation, the system defined by Eqs.~(\ref{eq:honplusoff}-\ref{eq:hoff}) and Eq.~\eqref{eq:master} describes an absorption refrigerator that can be switched on and off by changing the flux $\phi$ which in an experiment would correspond to a magnetic field. In particular, the $c$ oscillator is cooled below the temperature of the cold bath $T_c$ when $\phi=\pi/2$ and the fridge is in the ``on'' configuration (i.e. $\hat{H}=\hat{H}_{\rm on}$). In the ``off'' configuration ($\phi=0$ implying $\hat{H}=\hat{H}_{\rm off}$), the temperature of the $c$ oscillator is very close to $T_c$. This is illustrated in Fig.~\ref{fig:sketch} (b), where the temperature in the $c$ oscillator is plotted as a function of time when turning the refrigerator on and off periodically.

\section{Steady-state cooling}
We start by considering the refrigerator in the steady-state regime, i.e. $\partial_t\hat{\rho}=0$, and in the ``on'' configuration. Below we consider two quantities to characterize the refrigerator in the steady state. The achieved temperature in the $c$ oscillator and the coefficient of performance (COP) which quantifies how efficiently the heat from the hot reservoir is used to extract heat from the cold reservoir. We also derive a general condition for cooling based on the second law.

The main quantity of interest is the temperature achieved in the $c$ oscillator. Since the reduced steady state is not strictly thermal, we need to specify how to assign a temperature $\theta_c$ to oscillator $c$. Here we choose $\theta_c$ as the temperature of a thermal state with mean energy $\Omega_c\langle \hat{n}_c\rangle$. Since the thermal state maximizes the entropy for a given energy, the reduced state in oscillator $c$ has a strictly lower entropy than the thermal state it is compared to. We note that the reduced state is very close to a thermal state [cf.~App.~\ref{app:state}].

\begin{figure}[t!]
  \centering
  \includegraphics[width=\columnwidth]{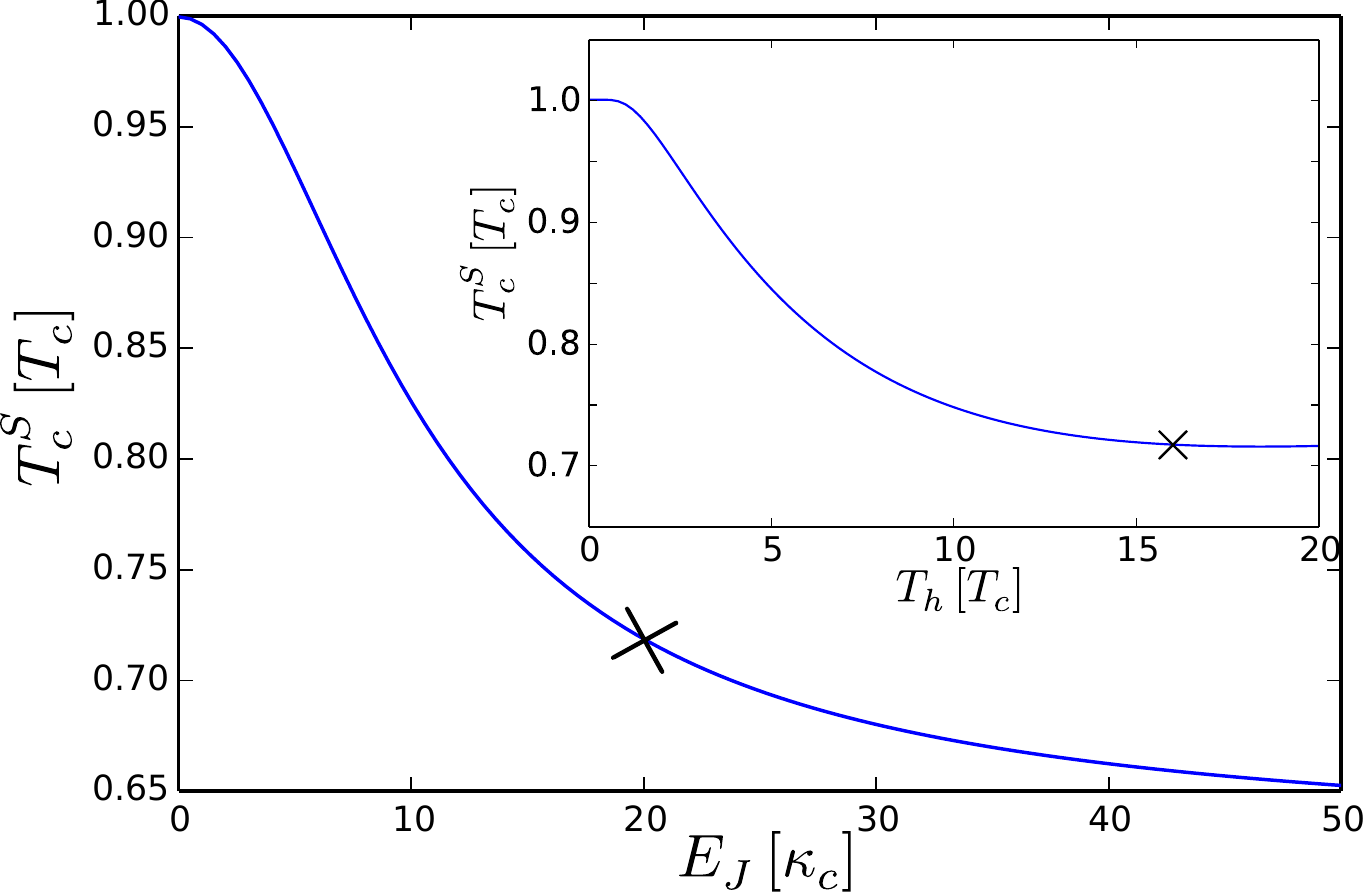}
  \caption{Performance of fridge as function of $E_J$. The higher the coupling between the oscillators, the stronger the cooling. Here we are limited by the condition $E_J\ll\Omega_\alpha$ which ensures the validity of our master equation. The inset shows the steady state temperature as a function of the hot temperature. While increasing $T_h$ generally enhances cooling, the non-linear operators given in Eq.~\eqref{eq:aops} reduce cooling for high temperatures $T_h$ because they only couple weakly to Fock states with a high photon number. The crosses show values as given in Tab.~\ref{tab:params}. All other parameters are as in Tab.~\ref{tab:params}.}
  \label{fig:steady}
\end{figure}

The steady-state temperature, denoted as $T_c^S$, is obtained numerically using the QuTiP library \cite{johansson:2013} and shown in Fig.~\ref{fig:steady} as a function of the Josephson energy and the temperature in the hot bath. Here we used a realistic set of parameters, given in Table \ref{tab:params}, based on recent experimental results \cite{hofheinz:2011,altimiras:2014,parlavecchio:thesis}. The refrigerator achieves $T_c^S\approx0.72T_c$, corresponding to cooling the $c$ oscillator from $50\,$mK to $36\,$mK. Moreover, we have verified numerically that the RWA resulting in Eqs.~(\ref{eq:honplusoff}-\ref{eq:hoff}) is a good approximation to the full Hamiltonian \eqref{eq:hamiltonian}, see App.~\ref{app:comp}.

In order to introduce the heat currents needed to obtain the COP, we consider the time-evolution of the mean photon number in the oscillators
\begin{equation}
\label{eq:ntder}
\partial_t\langle \hat{n}_\alpha\rangle = -i\langle[\hat{n}_\alpha,\hat{H}_{\rm on}]\rangle+\kappa_\alpha\left(n_B^\alpha-\langle \hat{n}_\alpha\rangle\right)=0.
\end{equation}
Here the first term corresponds to the change of photon number in oscillator $\alpha$ due to the interaction with the Josephson junction. This term crucially depends on the coherences between the two oscillators as can be seen by evaluating the commutator in the last expression. The unitary evolution thus exchanges photons between the oscillators one-by-one in a coherent fashion. The second term in Eq.~\eqref{eq:ntder} corresponds to photons being exchanged with the bath. We therefore define the average heat current as
\begin{equation}
\label{eq:heatcurr}
J_\alpha=\Omega_\alpha\kappa_\alpha (n_B^\alpha-\langle\hat{n}_\alpha\rangle),
\end{equation}
where the sign is chosen such that a positive heat flow indicates a flow from the bath to the oscillator. For the local master equation in Eq.~\ref{eq:master}, this heat current is equivalent to the definition $J_\alpha={\rm Tr}\{\hat{H}_{\rm on}\mathcal{L}_\alpha\hat{\rho}\}$, where $\mathcal{L}_\alpha$ denotes the superoperator responsible for the dissipation related to bath $\alpha$.

From Eq.~\eqref{eq:hon} and \eqref{eq:ntder}, we find (in the steady state)
\begin{equation}
\label{eq:heatcurrprop}
\frac{J_c}{\Omega_c}=\frac{J_h}{\Omega_h}=-\frac{J_r}{\Omega_r}.
\end{equation}
This proportionality is a consequence of the fact that for each photon removed from oscillator $c$, a single photon is removed from oscillator $h$ and added to oscillator $r$. Such a mechanism results in the universal COP
\begin{equation}
\label{eq:cop}
\eta=\frac{J_c}{J_h}=\frac{\Omega_c}{\Omega_h}.
\end{equation}

It can be shown that the last equation is bounded by the Carnot expression for the COP by considering the entropy change in the total system. The change in entropy in heat bath $\alpha$ is given by $\partial_tS_\alpha=-J_\alpha/T_\alpha$. Using the fact that the heat engine does not accumulate entropy in the steady state, the second law of thermodynamics implies that cooling ($J_c\geq0$) is obtained if
\begin{equation}
\label{eq:coolingcond}
\left(\frac{\Omega_r}{T_r}-\frac{\Omega_h}{T_h}-\frac{\Omega_c}{T_c}\right)\geq 0.
\end{equation}
This represent a general cooling condition. In turn, this implies
\begin{equation}
\label{eq:copbound}
\eta=\frac{\Omega_c}{\Omega_h}\leq \frac{1-\frac{T_r}{T_h}}{\frac{T_r}{T_c}-1}=\eta_C,
\end{equation}
where $\eta_C$ denotes the COP for a Carnot refrigerator \cite{skrzypczyk:2011}. The COP is thus bounded from above by $\eta_C$ which is reached in the reversible limit, where the steady-state tends to a tensor product of thermal states (at the respective bath temperatures) and the heat currents vanish.

\begin{table*}
\def\arraystretch{1.3}
\begin{tabular}{|c|c|c|c|c|c|c|c|c|c|c|c|}
\hline
  \,$\Omega_h/2\pi$\, & \,$\Omega_c/2\pi$\, & \,$\Omega_r/2\pi$\, & \,$\kappa_c/2\pi=\kappa_h/2\pi$\, &\,$\kappa_r/2\pi$\, &\, $E_J/2\pi$\, &\, $\lambda$ \,&
  \, $T_h$ \,&\, $T_c=T_r$\,&\, $T_c^S$ \\\hline\hline
     \,$4.5$\,GHZ\, &\, $1$\,GHZ\,&\, $5.5$\,GHZ\, &\, $0.01$\,GHz\, &\, $0.025$\,GHz\, & \,$0.2\,$GHz 
      \,& \,$0.3$\,&\, $768$\,mK\, & \,$50\,$mK\, & $36\,$mK\\
\hline
\end{tabular}
\caption{Realistic Parameters for operating the proposed refrigerator. Here $\lambda=\lambda_h=\lambda_c=\lambda_r$.}
\label{tab:params}
\end{table*}

\section{On/off cooling}
Having established the performance of the system as a refrigerator in the ``on'' mode, we now show that the refrigerator can be switched off by changing $\phi$, i.e. via the magnetic field. We stress that the only control that is needed in order to switch the refrigerator on and off is the external phase bias of the Josephson junction, i.e. no knowledge of the system parameters, such as temperatures and coupling constants, is required.

From Eq.~\eqref{eq:hoff}, we can anticipate that the cooling cannot be completely switched off since $\hat{H}_{\rm off}$ also induces cooling. However, in the ``off'' mode cooling happens by converting two $c$ photons plus two $h$ photons into two $r$ photons. This second order process has a prefactor of $(\lambda_c\lambda_h\lambda_r)^2$ which considerably reduces the cooling leading to a steady state temperature of $T_c^S\approx T_c$. In particular, for the parameters in Tab.~\ref{tab:params}, we obtain $T_c^S\approx
0.985 T_c$. The evolution of the temperature in the $c$ oscillator upon switching the refrigerator on and off is plotted in Fig.~\ref{fig:sketch} (b). 

To establish that the physics is indeed equivalent to a bosonic version of the small refrigerator discussed in the literature \cite{linden:2010prl,levy:2012}, we replace the non-linear $\hat{A}_\alpha(k)$ operators in the Hamiltonian with identity operators and a prefactor that is treated as a fitting parameter. This results in the simplified Hamiltonians 
\begin{subequations}
\begin{align}
\label{eq:toyon}
&\hat{H}_{\rm on}=-E'_J\left[\hat{a}_r^\dag\hat{a}_c\hat{a}_h+H.c.\right],\\
\label{eq:toyoff}
&\hat{H}_{\rm off}=E''_J\left[(\hat{a}_r^\dag)^2\hat{a}_c^2\hat{a}_h^2+H.c.\right].
\end{align}
\end{subequations}
This simplified model becomes exact in the limit $\lambda_\alpha\rightarrow0$ with $E_J'=8\lambda_c\lambda_h\lambda_rE_J$ and $E_J''=8\lambda_c^2\lambda_h^2\lambda_r^2E_J$ since there we have $\hat{A}_\alpha(k)=(2\lambda_\alpha)^k/k!$. As illustrated in Fig.~\ref{fig:sketch} (b), this model also captures the physics away from the $\lambda_\alpha\rightarrow0$ limit upon treating $E_J'$ and $E_J''$ as fitting parameters. The cooling is thus mediated by the three-body interaction in Eq.~\eqref{eq:toyon} and not the non-linear $\hat{A}_\alpha(k)$ operators.

\section{Transient cooling}
As previously discussed in the literature \cite{mitchison:2015,brask:2015,das:2016}, temperatures below the steady-state temperature can be obtained in the transient regime. The temperature in oscillator $c$ shows oscillations arising from the unitary evolution damped by the dissipative terms in the master equation. Depending on the parameters, these temperature oscillations go below the steady state temperature $T_c^S$. 

\begin{figure}[t!]
  \centering
  \includegraphics[width=\columnwidth]{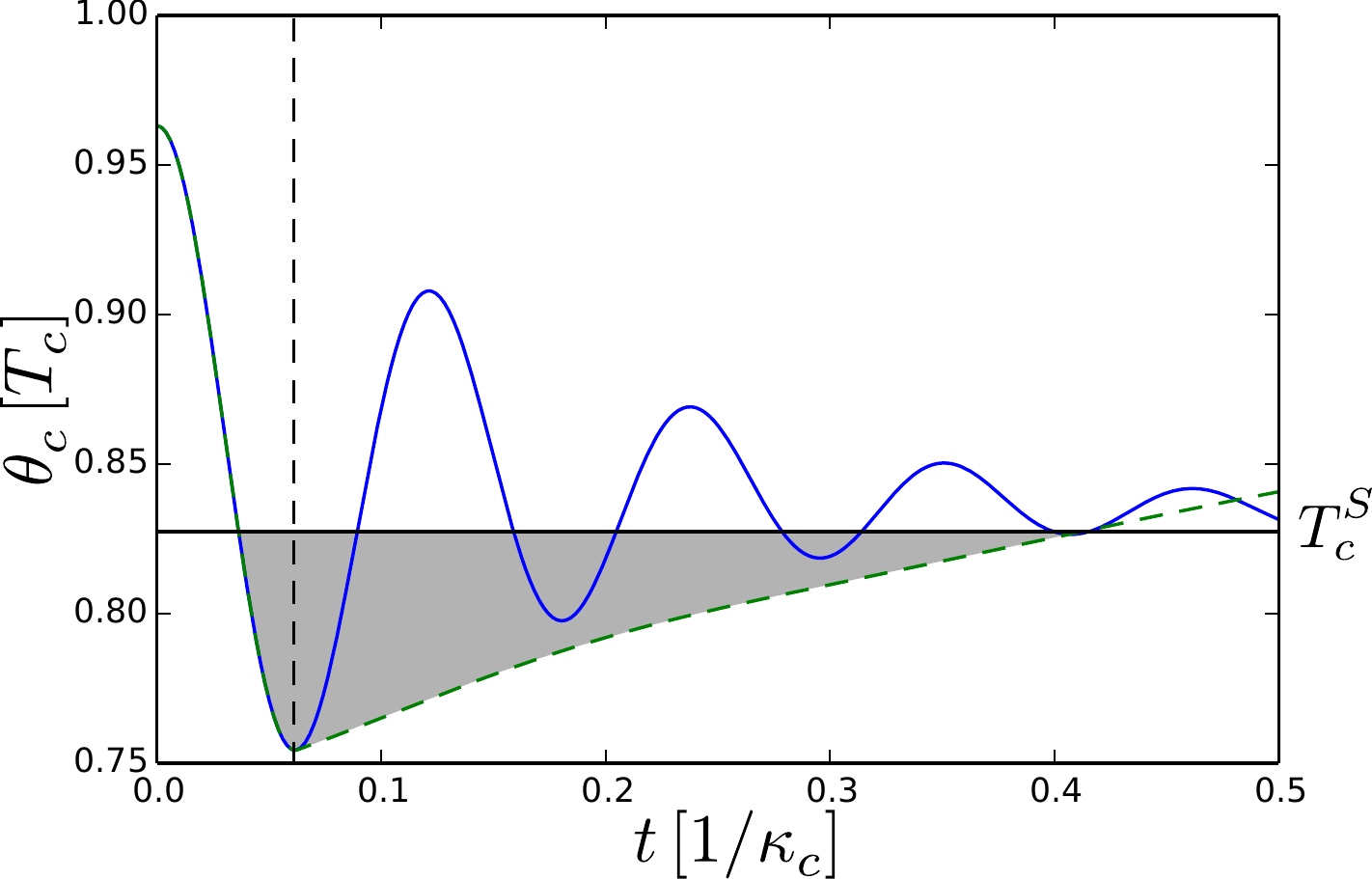}
  \caption{Transient cooling. Blue (solid) line: The temperature in the $c$ oscillator oscillates before reaching the steady state temperature $T_c^S$. Green (dashed) line: Switching off the fridge when the temperature reaches its first minimum allows for cooling below the steady state temperature. For low coupling to the cold bath, a temperature below $T_c^S$ can be maintained for a substantial amount of time (shaded area). Parameters are as in Tab.~\ref{tab:params} except for $\kappa_h=\kappa_c=\kappa_r=0.001\Omega_c$ and $k_BT_h=8\Omega_c$ which corresponds to half of the value in Tab.~\ref{tab:params}.}
  \label{fig:transient}
\end{figure}

In order to take advantage of this effect, it is however crucial to be able to switch the refrigerator on and off. Here we show that our model is tailored for this, and can thus benefit from coherence-enhanced cooling. This is demonstrated in Fig.~\ref{fig:transient}, where switching off the engine at the first minimum is shown to maintain the oscillator at a temperature below $T_c^S$ for a substantial amount of time. We note however that such a protocol requires precise timing and, depending on its implementation, might therefore no longer be fully autonomous. For the parameters in Tab.~\ref{tab:params}, only small signatures of the temperature oscillations are visible and they do not reach below $T_c^S$ [cf.~Fig.~\ref{fig:sketch}(b)]. In order to enhance the oscillations, one needs to decrease decoherence by reducing the couplings to the baths and/or the temperatures of the baths. 

\section{Feasibility and conclusion}
Using a set of realistic parameters in Tab.~\ref{tab:params}, we showed that a substantial cooling effect can be expected. The experimental prospects of these results will now be discussed. 
In Ref.~\cite{altimiras:2014}, a single oscillator with frequency $\sim\,$GHz was coupled to a normal tunnel junction, with coupling $\lambda\approx0.5$. For Josephson junctions, experiments on two oscillators (with GHz frequencies) with $\lambda\approx0.15$ have been performed and experiments on four oscillators are in preparation \cite{parlavecchio:thesis}. We are thus confident that coupling three oscillators with $\lambda\approx0.3$ is feasible. Note that we kept the couplings ($E_J$ and $\kappa_\alpha$) well below the frequencies in order to remain in the validity regime of our master equation, these parameters could be significantly increased in an experiment for testing different regimes. 

Another crucial ingredient for our proposal is the external phase bias which could be implemented using a magnetic field in a loop geometry [cf.~Fig.~\ref{fig:sketch} (a)], which is standard, e.g. in rf-SQUIDs. Finally, the harmonic oscillators need to be coupled to thermal baths at different temperatures. Specifically, the $h$ oscillator needs to be coupled to a bath at a temperature that is substantially higher than the temperature of the environment. Using a transmission line to feed the thermal noise from the hot bath to the $h$ oscillator would allow for a spatial separation of the hot bath and the rest of the setup.

In conclusion, we proposed an implementation for a quantum absorption refrigerator within reach of current technology. Moreover, an attractive feature of our model is a built-in on/off-switch, which allows one to take advantage of coherence-enhanced cooling. We hope that our study motivates further theoretical and experimental work on quantum thermal machines which provide a promising testbed to investigate the foundations of quantum thermodynamics.

\emph{Note added.} -- During the finishing of this manuscript, several related proposals appeared online \cite{marchegiani:2016,tan:2016,solinas:2016,karimi:2016}. While these works also propose quantum thermal machines based on the Josephson effect, they do not constitute absorption refrigerators. 

\acknowledgments
We acknowledge discussions with M. Woods, R. Uzdin, A. A. Clerk, J.-R. Souquet, and F. Portier. We acknowledge financial support from the Swiss National Science Foundation and QSIT.  MH further acknowledges funding by the Austrian Science Fund (FWF) through the START project Y879-N27, Swiss  National  Science  Foundation (AMBIZIONE Z00P2-161351). MPL acknowledges support from the Spanish MINECO (Project No. FIS2013-40627-P and FOQUS FIS2013-46768-P, Severo Ochoa grant SEV-2015-0522 and Grant No. FPU13/05988), Fundacion Cellex and the Generalitat de Catalunya (SGR875). We are grateful for support from the EU COST Action MP1209 “Thermodynamics in the quantum regime”.

\appendix

\section{Rotating wave approximation}
\label{app:rwa}
Here we discuss the derivation of Eqs.~(\ref{eq:honplusoff}-\ref{eq:hoff}) starting from the Hamiltonian in Eq.~\eqref{eq:hamiltonian}. The validity of the approximations that are made here are checked in App.~\ref{app:comp}.
Using the unitary transformation
\begin{equation}
\label{eq:unitrot}
\hat{U}=\prod_{\alpha=c,h,r}e^{i\hat{a}_\alpha^\dagger\hat{a}_\alpha\Omega_\alpha t},
\end{equation}
we transform the Hamiltonian into a rotating frame resulting in
\begin{widetext}
\begin{equation}
\label{eq:hamrot}
\hat{H}_R=\hat{U}^\dag \hat{H}\hat{U}+i\left(\partial_t\hat{U}^\dag\right)\hat{U}=-\frac{E_J}{2}e^{i\phi}\prod_{\alpha=c,h,r}\left[\sum\limits_{k=0}^{\infty}i^k(\hat{a}_\alpha^\dag)^k\hat{A}_\alpha(k)e^{ik\Omega_\alpha t}+\sum\limits_{k=1}^{\infty}i^k\hat{A}_\alpha(k)\hat{a}_\alpha^ke^{-ik\Omega_\alpha t}\right]+ H.c.
\end{equation}
where the operators $\hat{A}_\alpha(k)$ are given in Eq.~\eqref{eq:aops}. Making use of the resonance condition $\Omega_r=\Omega_h+\Omega_c$, we neglect all off-resonant terms. This results in the Hamiltonian
\begin{equation}
\label{eq:hamrotres}
\begin{aligned}
\hat{H}_{RWA}&=-\frac{E_J}{2}e^{i\phi}\sum_{k=0}^{\infty}(-i)^k\left[(\hat{a}_r^\dag)^k\hat{A}_c(k)\hat{A}_h(k)\hat{A}_r(k)\hat{a}_h^k\hat{a}_c^k+(\hat{a}_c^\dag)^k
(\hat{a}_h^\dag)^k\hat{A}_c(k)\hat{A}_h(k)\hat{A}_r(k)\hat{a}_r^k\right]+H.c.\\&=\sin\phi\hat{H}_{\rm on}+\cos\phi\hat{H}_{\rm off},
\end{aligned}
\end{equation}
with
\begin{equation}
\label{eq:honfull}
\hat{H}_{\rm on}=E_J\sum_{k=1}^{\infty}\left[(-1)^{k}(\hat{a}_r^\dag)^{2k-1}\hat{A}_c(2k-1)\hat{A}_h(2k-1)\hat{A}_r(2k-1)\hat{a}_h^{2k-1}\hat{a}_c^{2k-1}+H.c.\right],
\end{equation}
and
\begin{equation}
\label{eq:hofffull}
\hat{H}_{\rm off}=E_J\sum_{k=0}^{\infty}\left[(-1)^{k+1}(\hat{a}_r^\dag)^{2k}\hat{A}_c(2k)\hat{A}_h(2k)\hat{A}_r(2k)\hat{a}_h^{2k}\hat{a}_c^{2k}+H.c.\right].
\end{equation}
\end{widetext}
Equations (\ref{eq:honplusoff}-\ref{eq:hoff}) are then recovered by only keeping the $k=1$ terms in Eqs.~\eqref{eq:honfull} and \eqref{eq:hofffull}. We therefore neglect terms that are higher order in $\lambda_\alpha$. We note that in Eq.~\eqref{eq:hofffull} we also neglect a term which does not change the photon number in the oscillators (the $k=0$ term). The validity of all approximations made in this section is checked below.

\section{Comparison to full Hamiltonian}
\label{app:comp}
 \begin{figure}[h!]
   \centering
   \includegraphics[width=\columnwidth]{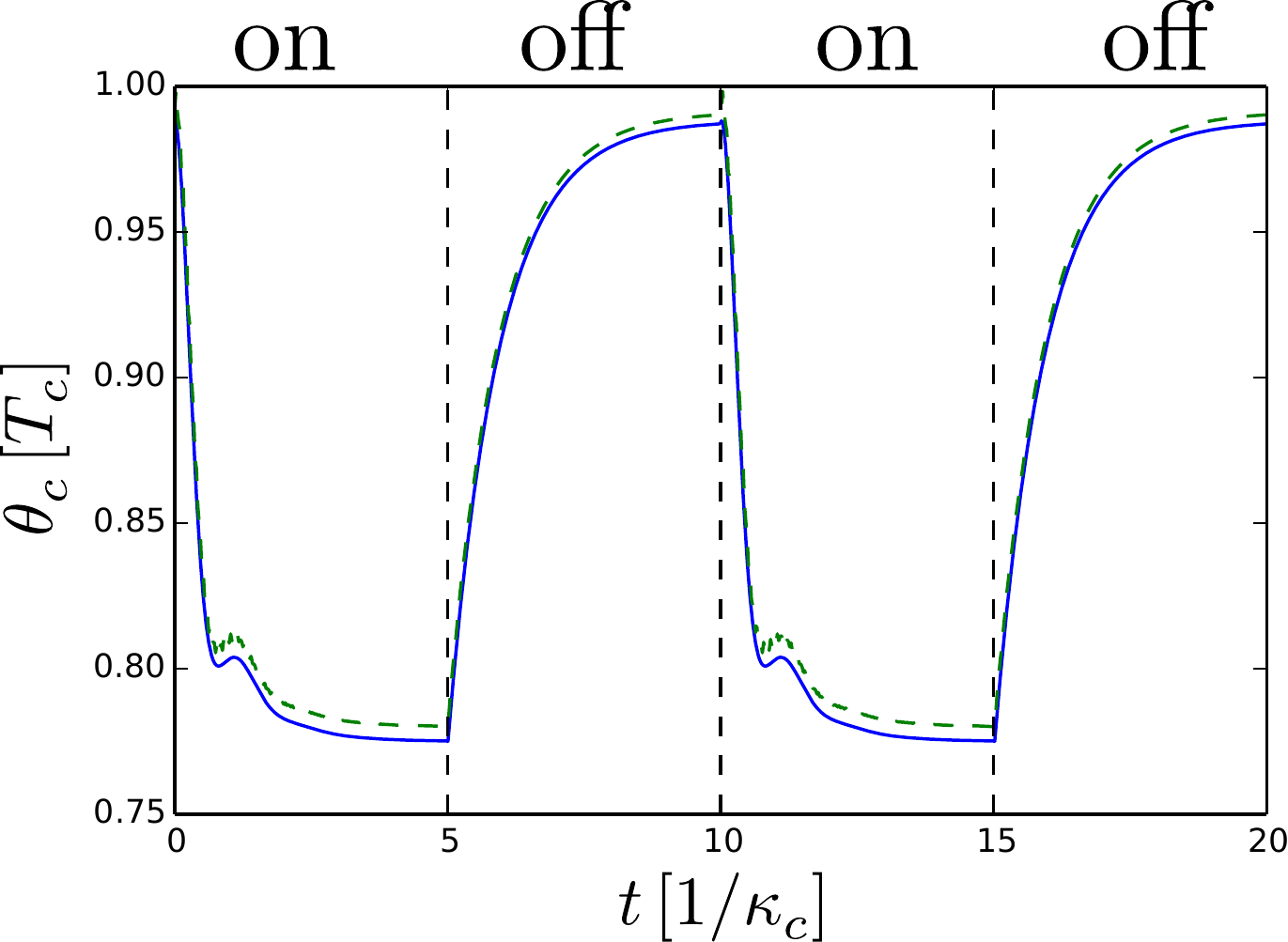}
   \caption{Temperature of the cold harmonic oscillator $\theta_c$ as a function of time. At the dashed line the fridge is switched on or off respectively. The solid (blue) line is obtained using Eqs.~\eqref{eq:hon} and \eqref{eq:hoff}, the dashed (green) line using the full Hamiltonian in Eq.~\eqref{eq:hamiltonian}. Apart from a small overestimation of the cooling power, the RWA approximation describes the system extremely well.}
   \label{fig:comparefull}
 \end{figure}

In this section, we compare the solutions obtained by Eqs.~(\ref{eq:honplusoff}-\ref{eq:hoff}) with a numerical solution of the full Hamiltonian in Eq.~\eqref{eq:hamiltonian}. Again we use the QuTiP library \cite{johansson2013qutip}, in particular the master equation solver. The results are shown in Fig.~\ref{fig:comparefull} and show excellent agreement between the full Hamiltonian and our approximations. This shows that neglecting non-resonant as well as well as higher order in $\lambda_\alpha$ terms is justified. Parameters are the same as in the main text (Tab.~\ref{tab:params}) with the exception of $T_h$ which is half the value used in the main text. This allows us to decrease the dimension of the Hilbert space sufficiently in order to numerically solve the full Hamiltonian.

\section{Reduced state in the cold oscillator}
\label{app:state}
In this section, we check that the reduced state in the $c$ oscillator, given by tracing out the degrees of freedom of the $h$ and $r$ oscillators, is close to a thermal state. 
 
To discuss the performance of the refrigerator one needs to assign a temperature to the oscillator being cooled. As the reduced state in the $c$ oscillator is in general not a thermal (Gibbs) state, there are a number of options to define temperature.
A natural approach is the following: if one has many copies of the cooled state and uses them to form a bath, what are the possible temperatures that this bath may have?

A first option is to assume no further manipulation of the systems. In the presence of any weak interaction between the systems, they will equilibrate to a state of maximum entropy, but of the same energy (via the first law). This corresponds to defining the temperature of the state via the thermal state of the same mean energy. This is the approach taken in the main text, which has the benefit of preserving the autonomous nature of the setup. Note that this is a conservative approach, in the sense that it gives the smallest estimate of the amount of cooling.

Another approach is to allow for arbitrary (hence non-autonomous in general) unitary operations on the oscillators. If the state is not \emph{completely passive} (i.e. Gibbs), one may find a unitary that extracts some energy. In the limit of a large number of systems, this unitary can extract the maximum allowed, i.e. it will leave each system in a thermal (passive) state of the same entropy as the original (since unitaries conserve entropy). This corresponds to defining the temperature via the thermal state of the same entropy. In general this leads to stronger estimated cooling than the first approach. 
 
\begin{figure}[b!]
   \centering
   \includegraphics[width=\columnwidth]{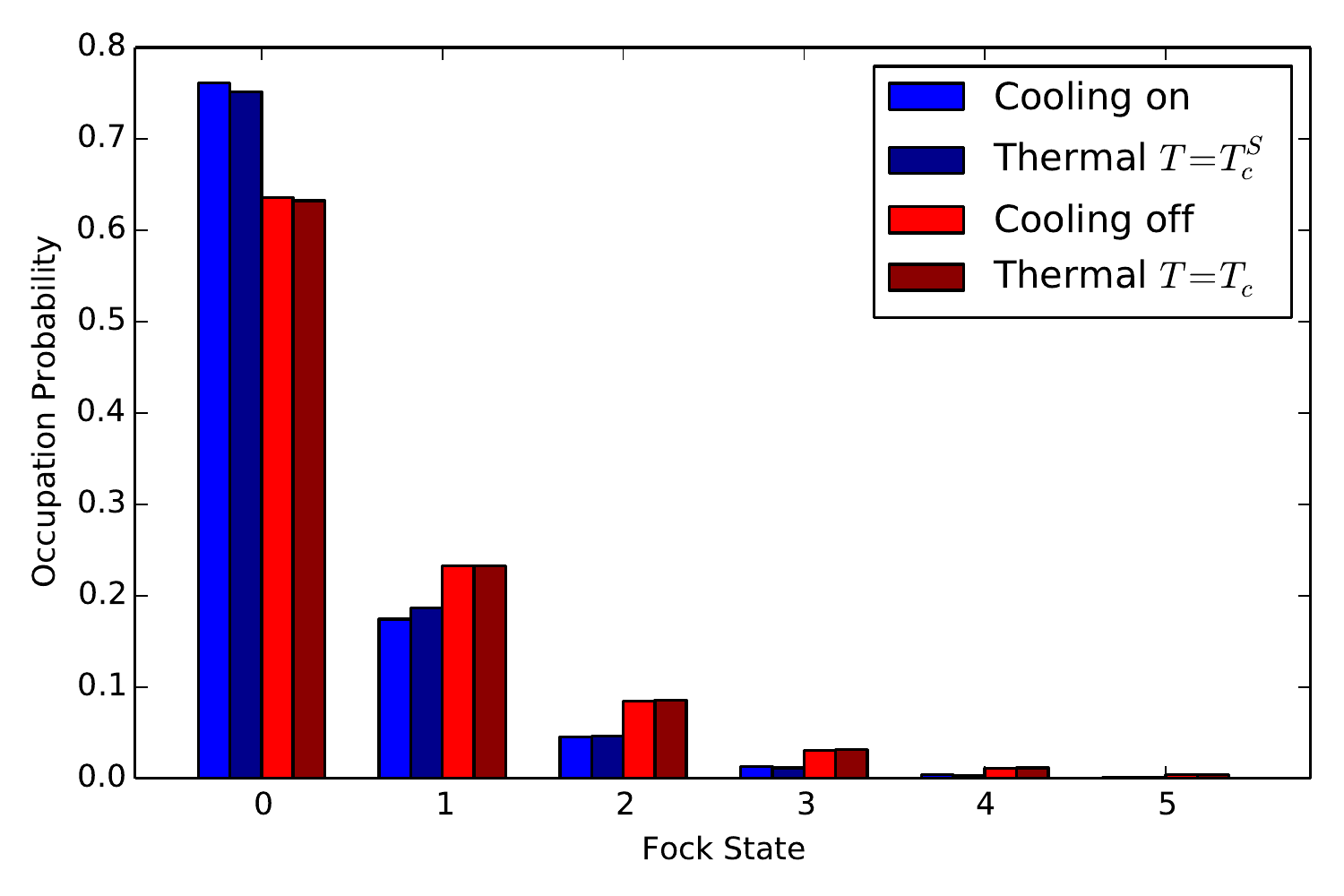}
   \caption{Fock state occupation probabilities. The light colors show the Fock state occupation probabilities for the reduced state in the $c$ oscillator if cooling is on (blue) and off (red) respectively. The dark colors show Fock state occupation probabilities for thermal states at the corresponding temperatures. The reduced states are very close to thermal states. Note that the reduced states are diagonal in the Fock state basis. Parameters are given in Tab.~\ref{tab:params}.}
   \label{fig:occupation}
 \end{figure}

In our case, both approaches lead to very similar temperature estimates, indicating that the reduced state in the $c$ oscillator is very close to a thermal state. This is confirmed in Fig.~\ref{fig:occupation}, where the Fock state occupation probabilities of the steady states in the ``on'' and in the ``off'' mode are compared to thermal states at the respective temperatures. The reduced states are shown to be extremely close to thermal states (note that the reduced states are diagonal in the Fock state basis).
 
\bibliography{biblio}

\end{document}